\documentclass{lmcs}
\pdfoutput=1

\usepackage{lastpage}
\lmcsdoi{15}{3}{3}
\lmcsheading{}{\pageref{LastPage}}{}{}%
{Jun.~14,~2018}{Jul.~24,~2019}{}


\usepackage{hyperref}

\def\ie{{\em i.e.}}

\usepackage{amssymb}
\usepackage{dsfont}
\usepackage{soul}

\newcommand{\ts}{\textstyle}

\newcommand{\mr}{\mathrm}

\newcommand{\mc}{\mathcal}

\newcommand{\uhr}{\upharpoonright}

\newcommand{\sq}[1]{\ensuremath{\langle#1\rangle}}

\newcommand{\eps}{\varepsilon}

\newcommand{\BAR}[1]{\overline{#1}}

\usepackage{mathtools}

\newcommand{\fieldfont}[1]{\mathbb{#1}}
\newcommand{\N}{\fieldfont{N}}

\newcommand{\F}{\fieldfont{F}}

\newcommand{\Span}{\mr{Span}}
\newcommand{\parity}{\textsc{parity}}

\newcommand{\Stab}{\mr{Stab}}

\newcommand{\ACzero}{\mr{AC}^0}
\newcommand{\size}{\mr{size}}
\newcommand{\leafsize}{\text{size}}
\newcommand{\leaf}{\mr{leafsize}}

\newtheorem*{restated}{Theorem 1.1 (restated)}

\begin{document}

\title[Subspace-Invariant AC$^{\text 0}$ Formulas]{Subspace-Invariant AC$^{\text 0}$ Formulas}
\titlecomment{A preliminary version of this paper appeared in ICALP 2017.}

\author[B.~Rossman]{Benjamin Rossman}
\address{University of Toronto}	
\email{ben.rossman@utoronto.ca} 
\thanks{Supported by NSERC and Sloan Research Fellowship.}

\begin{abstract}
We consider the action of a linear subspace $U$ of $\{0,1\}^n$ on the set of $\ACzero$ formulas with inputs labeled by literals in the set $\{X_1,\BAR X_1,\dots,X_n,\BAR X_n\}$, where an element $u \in U$ acts on formulas by transposing the $i$th pair of literals for all $i \in [n]$ such that $u_i=1$. A formula is {\em $U$-invariant} if it is fixed by this action. For example, there is a well-known recursive construction of depth $d+1$ formulas of size $O(n{\cdot}2^{dn^{1/d}})$ computing the $n$-variable \parity{} function; these formulas are easily seen to be $P$-invariant where $P$ is the subspace of even-weight elements of $\{0,1\}^n$. In this paper we establish a nearly matching $2^{d(n^{1/d}-1)}$ lower bound on the $P$-invariant depth $d+1$ formula size of \parity{}. Quantitatively this improves the best known $\Omega(2^{\frac{1}{84}d(n^{1/d}-1)})$ lower bound for {\em unrestricted} depth $d+1$ formulas \cite{rossman2015average}, while avoiding the use of the switching lemma. More generally, for any linear subspaces $U \subset V$, we show that if a Boolean function is $U$-invariant and non-constant over $V$, then its $U$-invariant depth $d+1$ formula size is at least $2^{d(m^{1/d}-1)}$ where $m$ is the minimum Hamming weight of a vector in $U^\bot \setminus V^\bot$.
\end{abstract}

\maketitle

\section{Introduction}\label{sec:intro}

There are two natural group actions on the set of literals $\{X_1,\BAR X_1,\dots,X_n,\BAR X_n\}$: the symmetric group $S_n$ acts by permuting indices, while $Z_2^n$ acts by toggling negations. 
These group actions extend to the set of $n$-variable Boolean functions, as well as the set of $n$-variable Boolean circuits. 
Here we consider bounded-depth circuits with unbounded fan-in AND and OR gates and inputs labeled by literals, also known as $\ACzero$ circuits.  
If $G$ is subgroup of $S_n$ or $Z_2^n$ (or more generally of the group $Z_2^n \rtimes S_n$ that they generate), we say that a function or circuit is {\em $G$-invariant} if it is fixed under the action of $G$ on the set of $n$-variable functions or circuits. Note that every $G$-invariant circuit computes a $G$-invariant function, and conversely every $G$-invariant function is computable by a $G$-invariant circuit. 

We define the {\em $G$-invariant circuit size} of a $G$-invariant function $f$ as the minimum number of gates in a $G$-invariant circuit that computes $f$. This may be compared to the {\em unrestricted circuit size} of $f$, noting that $f$ can be computed (possibly more efficiently) by circuits that are not $G$-invariant. Several questions arise. What gap, if any, exists between the $G$-invariant vs.\ unrestricted circuit size of $G$-invariant functions? Are lower bounds on $G$-invariant circuit size easier to obtain, and do they suggest new strategies for proving lower bounds for unrestricted circuits? Is there a nice characterization of functions computable by polynomial-size $G$-invariant circuits? The same questions may be asked with respect to $G$-invariant versions of other complexity measures, such as formula (leaf)size, as well as bounded-depth versions of both circuit and formula size, noting that the action of $G$ on circuits preserves both depth and fan-out.

The answer to these questions appears to be very different for subgroups of $S_n$ and subgroups of $Z_2^n$. This is illustrated by considering the $n$-variable parity function, which maps each element of $\{0,1\}^n$ to its Hamming weight modulo $2$. This function is both $S_n$-invariant (it is a so-called ``symmetric function'') and $P$-invariant where $P \subset Z_2^n$ is the index-2 subgroup of even-weight elements in $Z_2^n$. The smallest known circuits and formulas for $\parity_n$ have size $O(n)$ and leafsize $O(n^2)$, respectively. These circuits and formulas turn out to be $P$-invariant, as do the smallest known bounded-depth circuits and formulas (which we describe in \S\ref{sec:upper}). In contrast, the $S_n$-invariant circuit size of $\parity_n$ is known to be exponential \cite{anderson2016symmetric}.

\subsection{Invariance under subgroups of $S_n$}

$G$-invariant circuit complexity for subgroups $G$ of the symmetric group $S_n$ has been previously studied from the standpoint of Descriptive Complexity, an area of research concerned with the characterization of complexity classes in terms of definability in different logics \cite{immerman2012descriptive}.
Here one considers Boolean functions that encode isomorphism-invariant properties of relational structures.
Properties of $m$-vertex simple graphs, for instance, are identified with $G$-invariant functions $\{0,1\}^n \to \{0,1\}$ of $n = \binom{m}{2}$ variables, each corresponding to a potential edge, where $G$ is the group $S_m$ acting on the set of potential edges. More generally, if $\sigma$ is a finite relational signature $\sigma$, one considers the action of $S_m$ on $n = \sum_{R \in \sigma}m^{\mr{arity}(R)}$ variables encoding the possible $\sigma$-structures with universe $[m]$. 

Denenberg et al \cite{denenberg1986definability} showed that $S_m$-invariant circuits of polynomial size and constant depth (subject to a certain uniformity condition) capture precisely the first-order definable properties of finite $\sigma$-structures. Otto \cite{otto1996logic} introduced a certain limit object of finite circuits (imposing uniformity in a different way) 
and showed a correspondence between the logic $L^\omega_{\infty\omega}$ (infinitary logic with a bounded number of variables) and $S_m$-invariant circuits of polynomial size and arbitrary depth. Otto also gave characterizations of fixed-point logic and partial-fixed-point logic in terms of $S_m$-invariant Boolean networks. Recently, Anderson and Dawar \cite{anderson2016symmetric} showed a correspondence between fixed-point logic and polynomial-size $S_m$-invariant circuits, as well as between fixed-point logic with counting and polynomial-size $S_m$-invariant circuits in the basis that includes majority gates.

Choiceless Polynomial Time \cite{BGS99,BGS02,dawar2015symmetric,rossman2010choiceless} provides a different example of a $G$-invariant model of computation, where $G \subseteq S_n$ is the automorphism group of the input structure. Invariance under subgroups of $S_n$ has been explored in other settings as well, see for instance \cite{ajtai1994symmetric,riis2000generating,rudich1988optimal}.

\subsection{Invariance under subgroups of $Z_2^n$}

This paper initiates a study of invariant complexity with respect to subgroups of $Z_2^n$. Since our methods are linear algebraic, we shall henceforth identify $Z_2^n$ with the $\F_2$-vector space $\{0,1\}^n$ under coordinate-wise addition modulo $2$, denoted $\oplus$. We identify subgroups of $Z_2^n$ with linear subspaces $U$ of $\{0,1\}^n$. A function $f : \{0,1\}^n \to \{0,1\}$ is {\em $U$-invariant} if $f(x) = f(x\oplus u)$ for all $x \in \{0,1\}^n$ and $u \in U$. Note that $U$-invariant functions are in one-to-one correspondence with functions from the quotient space $\{0,1\}^n/U$ to $\{0,1\}$.

Our focus is on bounded-depth circuits and formulas. Returning to the example of the $P$-invariant function $\parity_n$ (where $P$ is the even-weight subgroup of $\{0,1\}^n$), there is a well-known recursive construction of depth $d+1$ circuits for $\parity_n$, which we describe in \S\ref{sec:upper}. Roughly speaking, one combines a depth 2 circuit for $\parity_{n^{1/d}}$ with depth $d$ circuits for $\parity_{n^{(d-1)/d}}$ on disjoint blocks of variables. This produces a depth $d+1$ circuit of size $O(n^{1/d}{\cdot}2^{n^{1/d}})$, which converts to a depth $d+1$ formula of leafsize $O(n{\cdot}2^{dn^{1/d}})$. Up to constant factors, these circuit and formulas are the smallest known computing $\parity_n$ and they are easily seen to be $P$-invariant, as we explain in \S\ref{sec:upper}.

The main result of this paper gives a nearly matching lower bound of $2^{d(n^{1/d}-1)}$ on the $P$-invariant depth $d+1$ formula size of $\parity{}_n$. This implies a $2^{n^{1/d}-1}$ lower bound on the $P$-invariant depth $d+1$ circuit size, via the basic fact that every ($U$-invariant) depth $d+1$ circuit of size $s$ is equivalent to a ($U$-invariant) depth $d+1$ formula of size at most $s^d$. Quantitatively, the lower bounds are stronger than the best known $\Omega(2^{\frac{1}{10}n^{1/d}})$ and $\Omega(2^{\frac{1}{84}d(n^{1/d}-1)})$ lower bounds for unrestricted depth $d+1$ circuits \cite{Hastad86} and formulas \cite{rossman2015average}, respectively. Of course, $P$-invariance is a severe restriction for circuits and formulas, so it is no surprise that the lower bounds we obtain is stronger and significantly easier to prove. The linear-algebraic technique in this paper is entirely different from the ``switching lemma'' approach of \cite{Hastad86,rossman2015average}. 

The general form of our lower bound is the following:

\begin{thm}\label{thm:main}
Let $U \subset V$ be linear subspaces of $\{0,1\}^n$, and suppose $F$ is a $U$-invariant depth $d+1$ formula which is non-constant over $V$. Then $F$ has size at least $2^{d(m^{1/d}-1)}$ where 
$m = \min\{|x| : x \in U^\bot \setminus V^\bot\}$, that is, the minimum Hamming weight of a vector $x$ which is orthogonal to $U$ and non-orthogonal to $V$.
\end{thm}

Here {\em size} refers to the number of depth 1 subformulas, as opposed to {\em leafsize}. Note that the bound in Theorem \ref{thm:main} does not depend on the dimension $n$ of the ambient space. Also note that aforementioned $2^{d(n^{1/d}-1)}$ lower bound for $\parity{}_n$ follows from the case $U = P$ and $V = \{0,1\}^n$. (Here $m = n$ is witnessed by the all-1 vector, which is an element of $P^\bot \setminus (\{0,1\}^n)^\bot$.)

We remark that, since $\lim_{d \to \infty} d(m^{1/d}-1) = \ln(m)$, Theorem \ref{thm:main} implies an $m^{\ln(2)}$ lower bound on the size of {\em unbounded-depth} formulas which are $U$-invariant and non-constant over $V$. Theorem \ref{thm:main} also implies a $2^{m^{1/d}-1}$ lower bound for depth $d+1$ circuits; however, we get no nontrivial lower bound for unbounded-depth circuits, since $\lim_{d \to \infty} m^{1/d}-1 = 0$.

\section{Preliminaries}\label{sec:prelims}

Let $n$ range over positive integers. $[n]$ is the set $\{1,\dots,n\}$. $\ln(n)$ is the natural logarithm and $\log(n)$ is the base-2 logarithm.

The Hamming weight of a vector $x \in \{0,1\}^n$, denoted $|x|$, is the cardinality of the set $\{i \in [n] : x_i = 1\}$. For vectors $x,y \in \{0,1\}^n$, let $x \oplus y$ denote the coordinate-wise sum modulo $2$ and let $\sq{x,y}$ denote the inner product modulo $2$.

Let $\mc L$ denote the lattice of linear subspaces of $\{0,1\}^n$. For $U,V \in \mc L$, let $U+V$ denote the subspace spanned by $U$ and $V$. Let $V^\bot$ denote the orthogonal complement $V^\bot = \{x \in \{0,1\}^n : \sq{x,v} = 0\text{ for all }v \in V\}$. We make use of the following facts about orthogonal complements over finite fields:
\begin{gather*}
\dim(V) + \dim(V^\bot) = n,\qquad
U \subseteq V \Longleftrightarrow V^\bot \subseteq U^\bot,\\
V = (V^\bot)^\bot,\qquad
(U + V)^\bot = U^\bot \cap V^\bot,\qquad
(U \cap V)^\bot = U^\bot + V^\bot.
\end{gather*}

\subsection{AC$^{\text 0}$ formulas}

We write $\mc F$ for the set of $n$-variable $\ACzero$ formulas (with unbounded fan-in AND and OR gates and leaves labeled by literals). Formally, let $\mc F = \bigcup_{d \in \N} \mc F_d$ where $\mc F_d$ is the set of {\em depth $d$ formulas}, defined inductively:
\begin{itemize}
\item $\mc F_0$ is the set $\{X_1,\BAR X_1,\dots,X_n,\BAR X_n\} \cup \{0,1\}$,
\item $\mc F_{d+1}$ is the set of ordered pairs
\begin{align*}\{(\mathsf{gate},\mc G) : \mathsf{gate} \in \{\mr{AND},\mr{OR}\} \text{ and } \mc G \text{ is a nonempty subset of }\mc F_d\}.\end{align*}
\end{itemize}

Every formula $F \in \mc F$ computes a Boolean function $\{0,1\}^n \to \{0,1\}$ in the usual way. For $x \in \{0,1\}^n$, we write $F(x)$ for the value of $F$ on $x$. For a nonempty set $S \subseteq \{0,1\}^n$ and $b \in \{0,1\}$, notation $F(S) \equiv b$ denotes that $F(x) = b$ for all $x \in S$. We say that $F$ is {\em non-constant} on $S$ if $F(S) \not\equiv 0$ and $F(S) \not\equiv 1$. 

The {\em depth} of $F$ is the unique $d \in \N$ such that $F \in \mc F_d$. {\em Leafsize} is the number of depth 0 subformulas, and {\em size} is the number of depth 1 subformulas. Inductively,
\begin{align*} 
  \leaf(F) &= 
  \begin{cases}
    \hspace{.8pt}1 &\text{if }F \in \mc F_0,\\
    \hspace{.8pt}\sum_{G \in \mc G} \leaf(G) &\text{if }F = (\mathsf{gate},\mc G) \in \mc F \setminus \mc F_0,
  \end{cases}\\
  \size(F) &= 
  \begin{cases}
    0 &\text{if }F \in \mc F_0,\\
    1 &\text{if }F \in \mc F_1,\\
    \sum_{G \in \mc G} \size(G)\phantom{\text{leaf}} &\text{if }F = (\mathsf{gate},\mc G) \in \mc F \setminus (\mc F_0 \cup \mc F_1).
  \end{cases}
\end{align*}
Clearly $\size(F) \le \leaf(F)$. Note that size is within a factor $2$ of the number of gates in $F$, which is how one usually measures the size of circuits. Our lower bound naturally applies to size, while the upper bound that we present in \S\ref{sec:upper} is naturally presented in terms of leafsize.

\subsection{The action of $\{0,1\}^n$}

We now formally define the action of $\{0,1\}^n$ (as the group $Z_2^n$) on the set $\mc F$. For $u \in \{0,1\}^n$ and $F \in \mc F$, let $F^u$ be the formula obtained from $F$ by exchanging literals $X_i$ and $\BAR X_i$ for every $i \in [n]$ with $u_i = 1$. Formally, this action is defined inductively by
\begin{align*}
  F^u = \begin{cases}
    F &\text{if $F \in \{0,1\}$,}\\
    X_i\text{ (resp.\ $\BAR X_i$)} &\text{if $F=X_i$ (resp.\ $\BAR X_i$) and $u_i = 0$,}\\
    \BAR X_i\text{ (resp.\ $X_i$)} &\text{if $F=X_i$ (resp.\ $\BAR X_i$) and $u_i = 1$,}\\
    (\mathsf{gate},\{G^u : G \in \mc G\}) &\text{if $F=(\mathsf{gate},\mc G)$.}
  \end{cases}
\end{align*}
Note that $F^u$ has the same depth and \leafsize{} as $F$ and computes the function $F^u(x) = F(x \oplus u)$ for all $x \in \{0,1\}^n$. 

Let $U$ be a linear subspace of $\{0,1\}^n$ (i.e.,\ subgroup of $Z_2^n$). We say that an $\ACzero$ formula $F$ is:
\begin{itemize}
  \item 
    {\em $U$-invariant} if $F^u = F$ (i.e.,\ these are syntactically identical formulas) for every $u \in U$,
  \item 
    {\em semantically $U$-invariant} if $F$ computes a $U$-invariant function (i.e.,\ $F(x) = F(x \oplus u)$ for every $u \in U$ and $x \in \{0,1\}^n$).
\end{itemize}
Note that every $U$-invariant formula is semantically $U$-invariant, but not conversely. For example, the formula $(\mr{AND},\{0,X_1,\dots,X_n\})$ computes the identically zero function and is therefore semantically $U$-invariant (for any $U$); however, this formula is not $U$-invariant (for any nontrivial $U$).

\subsection{Upper bound}\label{sec:upper}

We review the smallest known construction of bounded-depth formulas for $\parity_n$ (see \cite{Hastad86}) and observe that these formulas are $P$-invariant where $P$ is the even-weight subspace of $\{0,1\}^n$.

\begin{prop}\label{prop:upper}
For all $d,n \ge 1$, $\parity_n$ is computable by $P$-invariant depth $d+1$ formulas with either AND or OR as output gate and leafsize at most $n{\cdot}2^{dn^{1/d}}$.  
If $n^{1/d}$ is an integer, this bound improves to $n{\cdot}2^{d(n^{1/d}-1)}$.
\end{prop}

\proof 
Define $\beta(d,n)$ by the following recurrence:
\begin{align*}
  \beta(1,n) &= \begin{cases} 1 &\text{if }n=1,\\ \infty &\text{if }n > 1,\end{cases}
  \qquad
  \beta(d+1,n) = \min_{\substack{k,n_1,\dots,n_k \ge 1\,:\\ n_1+\dots+n_k=n}} 2^{k-1} \sum_{i=1}^k \beta(d,n_i).
\end{align*}
We will construct depth $d+1$ formulas of leafsize $\beta(d+1,n)$. If $n^{1/d}$ is an integer, we get the bound $\beta(d+1,n) \le n{\cdot}2^{d(n^{1/d}-1)}$ by setting $k=n^{1/d}$ and $n_1=\dots=n_k=n^{(d-1)/d}$. 
For arbitrary $d,n \ge 1$, we get the bound $\beta(d+1,n) \le n{\cdot}2^{dn^{1/d}}$ by setting $k = \lceil n^{1/d}\rceil$ and $n_1,\dots,n_k \in \{\lfloor n/k \rfloor, \lceil n/k \rceil\}$. 
In particular, note that $\beta(2,n) = n2^{n-1}$.

In the base case $d=1$, we have the brute-force DNF (OR-of-ANDs) and CNF (AND-of-ORs) formulas of leafsize $n2^{n-1}$ for \parity{}$_n$. These formulas are clearly $P$-invariant. Otherwise (if $d \ge 2$), fix the optimal choice of parameters $k,n_1,\dots,n_k$ for $\beta(d+1,n)$. Partition $[n]$ into sets $J_1 \sqcup \dots \sqcup J_k$ of size $|J_i| = n_i$. Let $\parity{}_{J_i}$ be the parity function over variables $\{X_j : j \in J_i\}$ and let $P_{J_i}$ be the subspace $\{u \in \{0,1\}^n : \bigoplus_{j \in J_i} u_j = 0\}$. 

By the induction hypothesis, for each $i \in [k]$ there exists a $P_{J_i}$-invariant formula $G_i$ computing $\parity{}_{J_i}$ with depth $d$ and leafsize at most $\beta(d,n_i)$ and output gate AND. Let $H_i$ be the formula obtained from $G_i$ by transposing literals $X_j$ and $\BAR X_j$ for any choice of $j \in J_i$; note that $H_i$ computes $1 - \parity{}_{J_i}$. Let $F$ be the brute-force DNF formula for \parity{}$_k$ over variables $Y_1,\dots,Y_k$.
We first form a depth $d+2$ formula $F'$ by replacing each literal $Y_i$ (resp.\ $\BAR Y_i$) in $F$ with the formula $G_i$ (resp.\ $H_i)$. The two layers of gates in $F'$ below the output consist entirely of AND gates; these two layers may be combined into a single layer, producing a formula $F''$ of depth $d+1$.
Since each variable $Y_i$ occurs in $2^{k-1}$ literals of $F$, the leafsize of $F''$ is $2^{k-1}\sum_{i=1}^k \beta(d,n_i)$ as required.
 
Finally, to see that $F''$ is $P$-invariant, consider an even-weight vector $u \in \{0,1\}^n$. Note that $u$ projects to an even-weight vector in $\{0,1\}^k$ whose $i$th coordinate is $\bigoplus_{j \in J_i} u_i$. Then $u$ acts on $F''$ by transposing subformulas $G_i$ and $H_i$ for all $i \in [k]$ such that $\bigoplus_{j \in J_i} u_i = 1$; therefore, $P$-invariance of $F''$ follows from $P_{\{Y_1,\dots,Y_k\}}$-invariance of $F$.
If we take $F$ to be a CNF instead of a DNF, the same construction produces $F''$ with OR instead of AND as its output gate.
\qed

\begin{rem}
$\parity{}_n$ is known to be computable by $P$-invariant formulas of depth $\lceil \log n \rceil + 1$ and leafsize $O(n^2)$ \cite{tarui2010smallest,Yab54}. The $n{\cdot}2^{dn^{1/d}}$ upper bound of Proposition \ref{prop:upper} is therefore slack, as this equals $n^3$ when $d = \log n$, whereas $n{\cdot}2^{d(n^{1/d}-1)} = n^2$. We suspect that the upper bound of Proposition \ref{prop:upper} can be improved that $O(n{\cdot}2^{d(n^{1/d}-1)})$ for all $d \le \log n$, perhaps by a more careful analysis of the recurrence for $\beta(d+1,n)$. Let us add that $\Omega(n^2)$ is a well-known lower bound for any depth, without the assumption of $P$-invariance \cite{khrapchenko1971complexity}.
\end{rem}

\section{Linear-algebraic lemmas}\label{sec:linear}

Recall that $\mc L$ denotes the lattice of linear subspaces of $\{0,1\}^n$. Let $U,V,S,T$ range over elements of $\mc L$. If $U$ is a subspace of $V$, recall that a {\em projection} from $V$ to $U$ is a linear map $\rho : V \to U$ such that $\rho(u)=u$ for every $u \in U$. We begin by showing that if $U$ is a codimension-$k$ subspace of $V$ (i.e.,\ $\dim(V) - \dim(U) = k$), then there there exists a projection $\rho : V \to U$ with ``Hamming-weight stretch'' $k+1$.

\begin{lem}\label{la:rho}
If $U$ is a codimension-$k$ subspace of $V$, then there exists a projection $\rho$ from $V$ to $U$ such that $|\rho(v)| \le (k+1)|v|$ for all $v \in V$.
\end{lem}

\proof 
Greedily choose a basis $w_1,\dots,w_k$ for $V$ over $U$ such that $w_i$ has minimal Hamming weight among elements of $V \setminus \Span(U\cup \{w_1,\dots,w_{i-1}\})$ for all $i \in [k]$. Each $v \in V$ has a unique representation $v = u \oplus a_1w_1 \oplus \dots \oplus a_kw_k$ where $u \in U$ and $a_1,\dots,a_k \in \{0,1\}$. Let $\rho : V \to U$ be the map $v \mapsto u$ and observe that this is a projection.

To show that $|\rho(v)| \le (k+1)|v|$, we first observe that $|a_iw_i| \le |v|$ for all $i \in [k]$. If $a_i = 0$, this is obvious, as $|a_iw_i| = 0$. If $a_i=1$, then $v \in V \setminus \Span(U\cup \{w_1,\dots,w_{i-1}\})$, so by our choice of $w_i$ we have $|a_iw_i| = |w_i| \le |v|$. Completing the proof, we have
\begin{align*}
  |\rho(v)| 
  &=
  |v \oplus a_1w_1 \oplus \dots \oplus a_kw_k|\\
  &\le
  |v| + |a_1w_1| + \dots + |a_kw_k|\\
  &\le
  (k+1)|v|.\qedhere
\end{align*}

\begin{defi}
Define sets $\mc L_2$ and $\mc L_4$ as follows:
\begin{align*}
  \mc L_2 &= \big\{(U,V) \in \mc L \times \mc L : 
  U \text{ is a codimension-1 subspace of } V
  \big\},\\
  \mc L_4 &= \big\{((S,T),(U,V)) \in \mc L_2 \times \mc L_2 : T \cap U = S\text{ and } T + U = V\big\}.\vphantom{t^{\ts|}}
\end{align*}
\end{defi}

The next lemma shows that $\mc L_4$ is anti-symmetric under orthogonal complementation.

\begin{lem}\label{la:duality}
For all $((S,T),(U,V)) \in \mc L_4$, we have $((V^\bot,U^\bot),(T^\bot,S^\bot)) \in \mc L_4$.
\end{lem}

\proof 
We use the properties of orthogonal complements stated in \S\ref{sec:prelims}. Consider any $((S,T),(U,V)) \in \mc L_4$. First note that $(V^\bot,U^\bot) \in \mc L_2$ by the fact that $U \subseteq V \Longrightarrow V^\bot \subseteq U^\bot$ and $\dim(U^\bot) - \dim(V^\bot) = (n-\dim(U)) - (n-\dim(V)) = \dim(V)-\dim(U) = 1$. Similarly, we have $(T^\bot,S^\bot) \in \mc L_2$. We now have $((V^\bot,U^\bot),(T^\bot,S^\bot)) \in \mc L_4$ since $U^\bot \cap T^\bot = (T + U)^\bot = V^\bot$ and $U^\bot + T^\bot = (T \cap U)^\bot = S^\bot$.
\qed

Finally, we state a dual pair of lemmas which play a key role in the proof of Theorem~\ref{thm:main}.

\begin{lem}\label{la:pre-main}
For all $(S,T) \in \mc L_2$ and $V \supseteq T$, there exists $U \supseteq S$ such that $((S,T),(U,V)) \in \mc L_4$ and
\begin{align*}
  \min_{x \in V \setminus U} |x| \ge \frac{1}{\dim(V)-\dim(T)+1}\min_{y \in T \setminus S}|y|.
\end{align*}
\end{lem}

\proof 
By Lemma~\ref{la:rho}, there exists a projection $\rho$ from $V$ onto $T$ such that $|\rho(v)| \le (\dim(V)-\dim(T)+1)|v|$ for all $v \in V$. Let $U = \rho^{-1}(S)$ and note that $U$ is a codimension-$1$ subspace of $V$. (This follows by applying the rank-nullity theorem to linear maps $\rho : V \to T$ and $\rho{\uhr}U : U \to S$ and noting that $\ker(\rho) = \ker(\rho{\uhr}U)$.) We have
$S = T \cap U$ and $T + U = V$, hence $((S,T),(U,V)) \in \mc L_4$. Choosing $x$ with minimum Hamming weight in $V \setminus U$, we observe that $\rho(x) \in T \setminus S$ and $|x| \ge |\rho(v)|/(\dim(V)-\dim(T)+1)$, which proves the lemma.
\qed

\begin{lem}\label{la:main} 
For all $(U,V) \in \mc L_2$ and $S \subseteq U$, there exists $T \subseteq V$ such that $((S,T),(U,V)) \in \mc L_4$ and
\begin{align*}
  \min_{x \in S^\bot \setminus T^\bot} |x| \ge 
  \frac{1}{\dim(U) - \dim(S) +1}
  \min_{y \in U^\bot \setminus V^\bot} |y|.
\end{align*}
\end{lem}

\proof 
Follows directly from Lemmas \ref{la:duality} and \ref{la:pre-main}.
\qed

\section{Proof of Theorem \ref{thm:main}}\label{sec:main}

We first prove the base case of Theorem \ref{thm:main} for depth $2$ formulas, also known as DNFs and CNFs. 

\begin{lem}\label{la:base-case}
Suppose $F$ is a depth $2$ formula and $(U,V) \in \mc L_2$ such that $F(U) \equiv b$ and $F(V \setminus U) \equiv 1-b$ for some $b \in \{0,1\}$. Then $\size(F) \ge 2^{m-1}$ and $\mr{leafsize}(F) \ge m{\cdot}2^{m-1}$ where $m = \min\{|x| : x \in U^\bot \setminus V^\bot\}$.
\end{lem}

Note that Lemma \ref{la:base-case} does not involve the assumption that $F$ is $U$-invariant.

\proof 
Assume that $F$ is a DNF formula (i.e.,\ an OR-of-ANDs formula) and $F(U) \equiv 0$ and $F(V \setminus U) \equiv 1$.  
This is without loss of generality: if $F$ were a DNF formula and $F(U) \equiv 1$ and $F(V \setminus U) \equiv 0$, then we may consider $F^w$ for any choice of $w \in V \setminus U$; this is a DNF formula of the same size and leafsize, but has $F^w(U) \equiv 0$ and $F^w(V \setminus U) \equiv 1$. The argument for CNF formulas is similar.

We may further assume that $F$ is minimal firstly with respect to the number of clauses and secondly with respect to the number of literals in each clause.

Consider any clause $G$ of $F$. This clause $G$ is the AND of some number $\ell$ of literals. Without loss of generality, suppose these literals involve the first $\ell$ coordinates. Let $\pi$ be the projection $\{0,1\}^n \to \{0,1\}^\ell$ onto the first $\ell$ coordinates. There is a unique element $p \in \{0,1\}^\ell$ such that $G(x) = 1 \Longleftrightarrow \pi(x) = p$ for all $x \in \{0,1\}^n$. Observe that $G(U) \equiv 0$ (since $F(U) \equiv 0$) and, therefore, $p \notin \pi(U)$.

We claim that $p \in \pi(V)$. To see why, assume for contradiction that $p \notin \pi(V)$. Then $G(V) \equiv 0$. But this means that the clause $G$ can be removed from $F$ and the resulting function $F'$ would still satisfy $F'(U) \equiv 0$ and $F'(V \setminus U) \equiv 1$, contradicting the minimality of $F$ with respect to number of clauses. 

For each $i \in [\ell]$, let $p^{(i)} \in \{0,1\}^\ell$ be the element obtained from $p$ by flipping its $i$th coordinate.
We claim that $p^{(1)},\dots,p^{(\ell)} \in \pi(U)$. Without loss of generality, we give the argument showing $p^{(\ell)} \in \pi(U)$. Let $G'$ be the AND of the first $\ell-1$ literals in $G$, and let $F'$ be the formula obtained from $F$ by replacing $G$ with $G'$. For all $x \in \{0,1\}^n$, we have $G(x) \le G'(x)$ and hence $F(x) \le F'(x)$. Therefore, $F'(V \setminus U) \equiv 1$. Now note that there exists $u \in U$ such that $F'(u) = 1$ (otherwise, we would have $F'(u) \equiv 0$, contradicting the minimality of $F$ with respect to the width of each clause). Since $F(u) = 0$ and $G'$ is the only clause of $F'$ distinct from the clauses of $F$, it follows that $G'(u) = 1$. This means that $u_{\{1,\dots,\ell-1\}} = p_{\{1,\dots,\ell-1\}}$. We now have $\pi(u) = p^{(\ell)}$ (otherwise, we would have $\pi(u) = p$ and therefore $G(u) = 1$ and $F(u) = 1$, contradicting that fact that $F(U) \equiv 0$).

Note that $p^{(1)},\dots,p^{(\ell)}$ span either the even-weight subspace of $\{0,1\}^\ell$ (if $p$ has odd weight) or all of $\{0,1\}^\ell$ (if $p$ has even weight). Since $p^{(1)},\dots,p^{(\ell)} \in \pi(U)$ and $p \in \pi(V) \setminus \pi(U)$, only the former is possible. That is, we have $\pi(V) = \{0,1\}^\ell$ and $\pi(U) = \{q \in \{0,1\}^\ell : |q|$ is even$\}$.
Therefore, $1^\ell \in \pi(U)^\bot \setminus \pi(V)^\bot$ (writing $1^\ell$ for the all-1 vector in $\{0,1\}^\ell$). It follows that $1^\ell0^{n-\ell} \in U^\bot \setminus V^\bot$ and, therefore, $\ell = |1^{\ell}0^{n-\ell}| \ge m$ (by definition of $m$).

We now observe that
\begin{align*}
  \Pr_{v \in V}[G(v)=1] = \Pr_{v \in V}[\pi(v)=p] = \Pr_{q \in \pi(V)}[q=p] = \Pr_{q \in \{0,1\}^\ell}[q=p] = 2^{-\ell} \le 2^{-m}.
\end{align*}
That is, each clause in $F$ has value $1$ over at most $2^{-m}$ fraction of points in $V$. Since the set $V \setminus U$ has density $1/2$ in $V$, we see that $2^{m-1}$ clauses are required to cover $V \setminus U$.

Subject to the stated minimality assumptions on $F$ (first with respect to the number of clauses and second to the width of each clause), we conclude that $F$ contains $\ge 2^{m-1}$ clauses, each of width $\ge m$. Therefore, $\size(F) \ge 2^{m-1}$ and $\mr{leafsize}(F) \ge m{\cdot}2^{m-1}$.
\qed

The induction step of Theorem \ref{thm:main} makes use of the following inequality. 

\begin{lem}\label{la:abc}
For all real $a,b,c \ge 1$, we have
$a + c(b/a)^{1/c} \ge (c+1)b^{1/(c+1)}$. This holds with equality iff $a = b^{1/(c+1)}$.
\end{lem}

\proof 
Taking the derivative of the lefthand side with respect to $a$, we get $\frac{\partial}{\partial a}\big(a+c(b/a)^{1/c}\big) = 1 - (b/a^{c+1})^{1/c}$. The function $a \mapsto a+c(b/a)^{1/c}$ is thus seen to have a unique minimum at $a = b^{1/(c+1)}$, where it takes value $(c+1)b^{1/(c+1)}$.
\qed

Onto the main result:

\begin{restated}
Let $U \subset V$ be linear subspaces of $\{0,1\}^n$, and suppose $F$ is a  $U$-invariant depth $d+1$ formula which is non-constant over $V$. Then $F$ has size at least $2^{d(m^{1/d}-1)}$ where $m = \min\{|x| : x \in U^\bot \setminus V^\bot\}$.
\end{restated}

\proof 
We first observe that it suffices to prove the theorem in the case where $(U,V) \in \mc L_2$, that is, $U$ has codimension-$1$ in $V$. To see why, note that for any $U \subset V$ such that $F$ is $U$-invariant and non-constant over $V$, there must exist $U \subset W \subseteq V$ such that $(U,W) \in \mc L_2$ and $F$ is non-constant over $W$. Assuming the theorem holds with respect to $U \subset W$, it also holds with respect to $U \subset V$, since $U^\bot \setminus W^\bot \subseteq U^\bot \setminus V^\bot$ and hence $\min\{|x| : x \in U^\bot \setminus W^\bot\} \ge \min\{|x| : x \in U^\bot \setminus V^\bot\}$.

Therefore, we assume $(U,V) \in \mc L_2$ and prove the theorem by induction on $d$. The base case $d=1$ is established by Lemma \ref{la:base-case}. For the induction step, let $d \ge 2$ and assume $F \in \mc F_{d+1}$ is a  $U$-invariant and non-constant over $V$. Without loss of generality, we consider the case where $F = (\mr{OR},\mc G)$ for some nonempty $\mc G \subseteq \mc F_d$. (The case where $F = (\mr{AND},\mc G)$ is symmetric, with the roles of $0$ and $1$ exchanged.)

Since $F$ is  $U$-invariant, we have $G^u \in \mc G$ for every $u \in U$ and $G \in \mc G$. We claim that it suffices to prove the theorem in the case where the action of $U$ on $\mc G$ is transitive (\ie $\mc G = \{G^u : u \in U\}$ for every $G \in \mc G$). To see why, consider the partition $\mc G = \mc G_1 \sqcup \dots \sqcup \mc G_t$, $t \ge 1$, into orbits under $U$. For each $i \in [t]$, let $F_i$ be the formula $(\mr{OR},\mc G_i)$. Note that $F_i$ is $U$-invariant and $U$ acts transitively on $\mc G_i$. Clearly, we have $F(v) = \bigvee_{i \in [t]} F_i(v)$ for all $v \in V$. Since every $U$-invariant Boolean function is constant over sets $U$ and $V \setminus U$ (using the fact that $U$ has codimension-$1$ in $V$), it follows that each $F_i$ satisfies either $F_i(V) \equiv 0$ or $F(v) = F_i(v)$ for all $v \in V$. (It cannot happen that $F_i(V) \equiv 1$ for any $i$, since that would imply $F(V) \equiv 1$.) Because $F$ is non-constant over $V$, it follows that there exists $i \in [t]$ such that $F(v) = F_i(v)$ for all $v \in V$. In particular, this $F_i$ is non-constant over $V$. Since $\size(F) \ge \size(F_i)$, we have reduced proving the theorem for $F$ to proving to theorem for $F_i$. 

In light of the preceding paragraph, we proceed under the assumption that $U$ acts transitively on $\mc G$. Fix an arbitrary choice of $G \in \mc G$. Let
\begin{align*}
  S &= \Stab_U(G)\ (= \{u \in U : G^u = G\}),\\ 
  a &= \dim(U) - \dim(S) + 1.
\end{align*}
By the orbit-stabilizer theorem, 
\begin{align*}
  |\mc G| = |\mr{Orbit}_U(G)| = [U : S] = |U|/|S| = 2^{a-1}.
\end{align*}
Since $\size(G') = \size(G)$ for every $G' \in \mc G$, we have
\begin{equation}\label{eq1}
  \size(F) 
  = \sum_{G' \in \mc G} \size(G') =
  |\mc G| \cdot \size(G)
  =
  2^{a-1} \cdot \size(G).
\end{equation}

We next observe that $G^u$ is  $S$-invariant for every $u \in U$ (in fact, $S = \Stab_U(G^u)$). This follows from the fact that $(G^u)^s = G^{u \oplus s} = (G^s)^u = G^u$ for every $s \in S$.

By Lemma \ref{la:main}, there exists $T$ such that $((S,T),(U,V)) \in \mc L_4$ and 
\begin{align*}
  \min_{x \in S^\bot \setminus T^\bot} |x|
  \ge 
  \frac{1}{\dim(U)-\dim(S)+1}\min_{y \in U^\bot \setminus V^\bot} |y| 
  = 
  \frac{m}{a}.
\end{align*}
We claim that there exists $u \in U$ such that $G^u$ is non-constant on $T$. There are two cases to consider:\bigskip

\noindent\underline{Case 1}:
{\em Suppose $F(U) \equiv 0$ and $F(V \setminus U) \equiv 1$.}\medskip

We have $G(U) \equiv 0$ and $G(V) \not\equiv 0$. Fix any $v \in V \setminus U$ such that $G(v) = 1$. In addition, fix any $w \in T \setminus U$ (noting that $T \setminus U$ is nonempty since $U + T = V$ and $U \subset V$). Let $u = v \oplus w$ and note that $u \in U$ (since $U$ is a codimension-$1$ subspace of $V$ and $v,w \in V \setminus U$). We have $G^u(U) \equiv 0$ and $G^u(w) = G(w \oplus u) = G(v) = 1$. By the $S$-invariance of $G^u$, it follows that $G^u(S) \equiv 0$ and $G^u(T \setminus S) \equiv 1$. In particular, $G^u$ is non-constant on $T$.\bigskip

\noindent\underline{Case 2}:
{\em Suppose $F(U) \equiv 1$ and $F(V \setminus U) \equiv 0$.}\medskip

We have $G(U) \not\equiv 0$ and $G(V \setminus U) \equiv 0$. Fix any $u \in U$ such that $G(u) = 1$. In addition, fix any $w \in T \setminus U$ and let $v = w \oplus u$. We have $G^u(v) = G(v \oplus u) = G(w) = 0$ (since $w \in V \setminus U$ and $G(V \setminus U) \equiv 0$). We also have $G^u(\vec 0) = G(u) = 1$ where $\vec 0$ is the origin in $\{0,1\}^n$. By  $S$-invariance of $G^u$, it follows that $G^u(S) \equiv 1$ and $G^u(T \setminus S) \equiv 0$. In particular, $G^u$ is non-constant on $T$.\bigskip

Since $G^u$ is  $S$-invariant and non-constant on $T$ and $\mr{depth}(G^u) = (d-1)+1$, we may apply the induction hypothesis to $G^u$. Thus, we have
\begin{equation}\label{eq2}
  \size(G) = \size(G^u) \ge 2^{(d-1)((m/a)^{1/(d-1)}-1)}.
\end{equation}
Since $d \ge 2$, Lemma \ref{la:abc} tells us
\begin{equation}\label{eq3}
  a + (d-1)(m/a)^{1/(d-1)} \ge d(m/a)^{1/d}.
\end{equation}
Putting together (\ref{eq1}), (\ref{eq2}), (\ref{eq3}), we get the desired bound
\begin{align*}
  \size(F)
  &\ge 
  2^{a-1} \cdot 2^{(d-1)((m/a)^{1/(d-1)}-1)}\\
  &= 
  2^{a + (d-1)(m/a)^{1/(d-1)} - d}\\
  &\ge
  2^{d(m^{1/d}-1)}.
\end{align*}
This completes the proof of Theorem \ref{thm:main}.\qed

\section{Remarks and open questions}\label{sec:conclusion}

\subsection{Another application of Theorem \ref{thm:main}}

Theorem \ref{thm:main} applies to interesting subspaces $U$ of $\{0,1\}^n$ besides the even-weight subspace $P$. Here we describe one example. Let $G$ be a simple graph with $n$ edges, so that $\{0,1\}^n$ may be identified with the set of spanning subgraphs of $G$. The {\em cycle space} of $G$ is the subspace $Z \subseteq \{0,1\}^n$ consisting of {\em even subgraphs} of $G$ (i.e.,\ spanning subgraphs in which every vertex has even degree). Consider the even-weight subspace $Z_0 = \{z \in Z : |z|$ is even$\}$. Provided that $G$ is non-bipartite, $Z_0$ is a codimension-1 subspace of $Z$. 

Let $m = \min\{|x| : x \in Z_0^\bot \setminus Z^\bot\}$ as in Theorem \ref{thm:main} with $U = Z_0$ and $V = Z$. This number $m$ is seen to be equal to the minimum number of edges whose removal makes $G$ bipartite. It follows that $m = n - c$ where $c$ is the number edges in a maximum cut in $G$. Now suppose $G$ is generated as a uniform random $3$-regular graph with $n$ edges (and $\frac23 n$ vertices). There is a constant $\eps > 0$ such that $c \le (1-\eps)n$ (and hence $m \ge \eps n$) holds asymptotically almost surely \cite{bollobas1988isoperimetric}. From these observations, we have

\begin{cor}\label{cor:main}
Every $Z_0$-invariant depth $d+1$ formula that computes $\parity{}_n$ over $Z$ has size at least $2^{d((\eps n)^{1/d}-1)}$ asymptotically almost surely.
\end{cor}

The $\ACzero$ complexity of computing $\parity{}_n$ over the cycle space of a graph $G$ is loosely related to the $\ACzero$-Frege proof complexity of the Tseitin tautology on $G$, which has been explored recently in \cite{haastad2017small,pitassi2016poly}. In general, however, we do not have techniques to lower bound the (non-subspace-invariant) $\ACzero$ complexity of $\parity{}_n$ over arbitrary subspaces of $\{0,1\}^n$.

\subsection{The $V \setminus U$ search problem}\label{sec:game}

For linear subspaces $U \subset V$ of $\{0,1\}^n$, consider the following ``$V \setminus U$ search problem''. There is a hidden vector $w \in V \setminus U$ and the goal is to learn a nonzero coordinate of $w$ (any $i \in [n]$ such that $w_i = 1$) by asking queries (yes/no questions) in the form of linear functions $\{0,1\}^n \to \{0,1\}$. The {\em $d$-round query complexity} of this problem is the minimum number of queries required by a deterministic protocol which issues batches of queries over $d$ consecutive rounds. 
By an argument similar to the proof of Theorem \ref{thm:main}, we get a $d(m^{1/d}-1)$ lower bound on the $d$-round query complexity of the $V \setminus U$-search problem where $m = \min\{|x| : x \in U^\bot \setminus V^\bot\}$.  We remark that this $V \setminus U$ search problem may be viewed as an $U$-invariant version of the Karchmer-Wigderson game.

\subsection{Open questions}\label{sec:open}

We conclude by mentioning some open questions and challenges raised by this work:

\begin{itemize}
\item
Does the $2^{d(m^{1/d}-1)}$ lower bound of Theorem \ref{thm:main} (or even a weaker bound like $2^{\Omega(m^{1/d})}$ or $2^{m^{\Omega(1/d)}}$) apply to depth $d+1$ formulas which are {\em semantically $U$-invariant} and non-constant on $V$?
\item
Counting leafsize instead of size, improve the lower bound of Theorem \ref{thm:main} from $2^{d(m^{1/d}-1)}$ to $m{\cdot}2^{d(m^{1/d}-1)}$.
\item
Improve the upper bound of Proposition \ref{prop:upper} from $n{\cdot}2^{dn^{1/d}}$ to $O(n{\cdot}2^{d(n^{1/d}-1)})$ for all $d \le \log n$.
\item
What is the maximum gap, if any, between the $U$-invariant vs.\ unrestricted $\ACzero$ complexity of a $U$-invariant Boolean function?
\end{itemize}

\bibliographystyle{alpha}

\end{document}